\documentclass[sigconf]{acmart}
\newif\ifecomment
\ecommenttrue
\ifecomment 
\newcounter{ECommentNum}
\stepcounter{ECommentNum}
\newcommand{\enote}[2]{({\color{blue}{#1:}\arabic{ECommentNum}\stepcounter{ECommentNum}: #2})}
\else
\newcommand{\enote}[2]{}
\fi

\newenvironment{itemize-s}%
{\begin{itemize}%
		\setlength{\itemsep}{0pt}%
		\setlength{\parskip}{0pt}}%
	{\end{itemize}}

\renewcommand\footnotetextcopyrightpermission[1]{} % removes conference notes
\usepackage[english]{babel}
\usepackage[utf8]{inputenc}
\usepackage[colorinlistoftodos]{todonotes}
\usepackage{epsfig,url}
\usepackage{subcaption}
\usepackage{balance}
\usepackage{adjustbox}
\usepackage{epstopdf}
\usetikzlibrary{positioning,calc}
\usepackage{longtable}
\usepackage{siunitx}
\usepackage{enumerate}
\usepackage{enumitem}
\usepackage{rotating}
\usepackage{hhline}
\usepackage[printonlyused]{acronym}
\usepackage{tabularx}
\usepackage{booktabs}
\usepackage{xspace}
\usepackage{array}
\newcolumntype{x}[1]{>{\centering\arraybackslash\hspace{0pt}}p{#1}}
\usepackage[yyyymmdd,hhmmss]{datetime}
\usepackage{balance}

\acrodef{IRB}{Institutional Review Board}
\acrodef{SAI}{Send Authentication Information}
\acrodef{UL}{Update Location}
\acrodef{SoR}{Steering of Roaming}
\acrodef{AS}{Autonomous System}
\acrodef{DPA}{Diameter Proxy Agent}
\acrodef{DRA}{Diameter Routing Agent}
\acrodef{STP}{Signaling Transfer Point}
\acrodef{HLR}{Home Location Registry}
\acrodef{VLR}{Visiting Location Registry}
\acrodef{IPX-P}{IPX Provider}
\acrodef{ASP}{Application Service Provider}
\acrodef{HSS}{Home Subscriber Server}
\acrodef{MVNE}{Mobile Virtual Network Enabler}
\acrodef{VPN}{Virtual Private Network}
\acrodef{DEA}{Diameter Edge Agent}
\acrodef{RCS}{Rich Communication Services}
\acrodef{IMS}{IP Multimedia Subsystem}
\acrodef{MPLS}{Multiprotocol Label Switching}
\acrodef{MMS}{Multimedia Messaging Service}
\acrodef{SMS}{Short Messaging Service}
\acrodef{ENUM}{E.164 Number translation service}
\acrodef{IGP}{International Gateway Providers}
\acrodef{SS7}{Signaling System No. 7}
\acrodef{MAP}{Mobile Application Protocol}
\acrodef{GRX}{GPRS Roaming eXchange Service}
\acrodef{BSC}{Base Station Controller}
\acrodef{EPC}{Evolved Packet Core}
\acrodef{PDP}{Packet Data Protocol}
\acrodef{GMM}{GPRS Mobility Management}
\acrodef{GGSN}{Gateway GPRS Support Node}
\acrodef{NAS}{Non-Access Stratum}
\acrodef{SIGTRAN}{Signaling Transport}
\acrodef{RAN}{Radio Access Network}
\acrodef{GPRS}{General Packet Radio System}
\acrodef{UTRAN}{UMTS Terrestrial Radio Access Network}
\acrodef{SGSN}{Serving GPRS Support Node}
\acrodef{CN}{Core Network}
\acrodef{RNC}{Radio Network Controller}
\acrodef{MSC}{Mobile Switching Center}
\acrodef{MME}{Mobility Management Entity}
\acrodef{UMTS}{Universal Mobile Telecommunication Systems}
\acrodef{xDR}{eXtended Detail Record}
\acrodef{TAC}{Type Allocation Code}
\acrodef{PoP}{Point of Presence}
\acrodef{IMEI}{International Mobile Equipment Identity}
\acrodef{EMSISDN}{Encrypted Mobile Station International Subscriber Directory Number}
\acrodef{MSISDN}{Mobile Station International Subscriber Directory Number}
\acrodef{MME}{Mobility Management Entity}
\acrodef{VoLTE}{Voice over \ac{LTE}}
\acrodef{MVNO}{Mobile Virtual Network Operators}
\acrodef{CDR}{Call Detail Record}
\acrodef{IMSI}{International Mobile Subscriber Identity}
\acrodef{RAT}{Radio Access Technology}
\acrodef{LPWA}{Low Power Wide Area}
\acrodef{VMNO}{Visited Mobile Network Operator}
\acrodef{HMNO}{Home Mobile Network Operator}
\acrodef{DCH}{Data Clearing House}
\acrodef{GSMA}{GSM Association}
\acrodef{GSM}{Global System for Mobile communications}
\acrodef{TAP}{Transferred Account Procedure}
\acrodef{LTE-M}{\ac{LTE} Machine Type Communication}
\acrodef{MTC}{Machine Type Communications}
\acrodef{NB-IoT}{Narrow Band \ac{IoT}}
\acrodef{IOT}{Inter Operator Tariff}
\acrodef{IoT}{Internet of Things}
\acrodef{M2M}{Machine-to-Machine}
\acrodef{FNO}{Fixed Network Operator}
\acrodef{SP}{Service Provider} 
\acrodef{ASP}{Application Service Providers}
\acrodef{IPX}{IP Packet Exchange}
\acrodef{2G}{Second Generation}
\acrodef{3G}{Third Generation}
\acrodef{4G}{Fourth Generation}
\acrodef{5G}{Firth Generation}
\acrodef{ADB}{Android Debug Bridge}
\acrodef{ASCI}{Advertising Standards Council of India}
\acrodef{ASN}{Autonomous System Number}
\acrodef{CDN}{Content Delivery Network}
\acrodef{DL}{Downlink}
\acrodef{DNS}{Domain Name Service}
\acrodef{EaaS}{Experiment as a Service}
\acrodef{ECDF}{Empirical Cumulative Distribution Function}
\acrodef{HTTP}{Hyper Text Transfer Protocol}
\acrodef{ICMP}{Internet Control Message Protocol}
\acrodef{ISP}{Internet Service Provider}
\acrodef{IXP}{Internet Exchange Point}
\acrodef{LTE}{Long Term Evolution}
\acrodef{MBB}{Mobile Broadband}
\acrodef{E2E}{End-to-end}
\acrodef{QCI}{QoS Class Identifier}
\acrodef{NDT}{Network Diagnostic Tool}
\acrodef{QoE}{Quality of Experience}
\acrodef{QoS}{Quality of Service}
\acrodef{OS}{Operating System}
\acrodef{APN}{Access Point Name}
\acrodef{RMBT}{RTR Multithreaded Broadband Test}
\acrodef{RTT}{Round-Trip Time}
\acrodef{SIM}{Subscriber Identity Module}
\acrodef{SIMs}{Subscriber Identity Modules}
\acrodef{SLA}{Service-Level Agreement}
\acrodef{TCP}{Transmission Control Protocol}
\acrodef{UDP}{User Datagram Protocol}
\acrodef{UE}{User Equipment}
\acrodef{NRA}{National Regulatory Authority}
\acrodef{EC}{European Commission}
\acrodef{SGW}{Serving Gateway}
\acrodef{PGW}{Packet Data Network Gateway}
\acrodef{GTP}{GPRS Tunneling Protocol}
\acrodef{MNO}{Mobile Network Operator}
\acrodef{EU}{European Union}
\acrodef{HR}{home-routed roaming}
\acrodef{LBO}{local breakout}
\acrodef{IHBO}{IPX hub breakout}
\acrodef{VoIP}{Voice over IP}
\acrodef{IPG}{Inter-Packet Gap}
\acrodef{KS}{Kolmogorov-Smirnov}
\acrodef{FQDN}{Fully Qualified Domain Name}
\acrodef{MNC}{Mobile Network Code}
\acrodef{MCC}{Mobile Country Code}
\acrodef{MCCMNC}{\ac{MCC}-\ac{MNC}}
\acrodef{SIP}{Session Initiation Protocol}

\setcopyright{none}
\renewcommand\footnotetextcopyrightpermission[1]{}
\acmDOI{}
\acmISBN{}

\acmPrice{}
\hypersetup{draft} 	

\begin{document}
\title{A first look at the IP eXchange Ecosystem}

\author{Andra Lutu}
\affiliation{
	\institution{Telefonica Research}
}
\email{andra.lutu@telefonica.com}

\author{Byunjin Jun}
\affiliation{
	\institution{Northwestern University}
}
\email{byungjinjun2022@u.northwestern.edu}

\author{Fabian Bustamante}
\affiliation{
	\institution{Northwestern University}
}
\email{fabianb@cs.northwestern.edu}

\author{Diego Perino}
\affiliation{
	\institution{Telefonica Research}
}
\email{diego.perino@telefonica.com}

\author{Marcelo Bagnulo}
\affiliation{
	\institution{University Carlos III of Madrid}
}
\email{marcelo@it.uc3m.es}

\author{Carlos Gamboa Bontje }
\affiliation{
	\institution{Telefonica}
}
\email{carlos.gamboabontje@telefonica.com}

%\renewcommand{\shortauthors}{X.et al.}

\begin{abstract}
The IPX Network interconnects about 800 \acp{MNO} worldwide and a range of other service providers (such as cloud and content providers). It forms the core that enables global data roaming while supporting emerging applications, from VoLTE and video streaming to IoT verticals. This paper presents the first characterization of this, so-far opaque, IPX ecosystem and a first-of-its-kind in-depth analysis of ann \ac{IPX-P}.  
The IPX Network is a private network formed by a small set of tightly interconnected \acp{IPX-P}. We analyze an operational dataset from a large \ac{IPX-P} that includes BGP data as well as statistics from signaling. We shed light on the structure of the IPX Network as well as on the temporal, structural and geographic features of the IPX traffic. Our results are a first step in understanding the IPX Network at its core, key to fully understand the global mobile Internet.
\end{abstract}

\maketitle

% !TeX root = sigcomm20.tex
% ================================================================

\section{Introduction}
\label{sec:introduction}

International roaming is an important feature of cellular networks, allowing subscribers to use their 
devices anywhere in the world as if at home. 
Under the \ac{IPX} model~\cite{ipx-gsm-whitepaper, ir34}, \acp{MNO} contract the services of third party providers -- the \acf{IPX-P} -- to offer their customers access to mobile services in any foreign country.
No \ac{IPX-P} on its own is able to provide connections on a global basis (e.g., single-handily
interworking with all \acp{MNO}). Thus, \acp{IPX-P} peer to other \acp{IPX-P}~\cite{apnic17peering} to expand their geographical footprint. 
The resulting  \textit{IPX Network}, is an isolated network that bypases the public
Internet~\cite{manynet}, to ensure secure, SLA-compliant services, from video streaming and AR/VR to IoT verticals, such
as connected cars. 

Recent years have brought a rapid growth in the number of participants in the IPX ecosystem and the volume of traffic 
they exchanged. The growing number of international travelers, reaching 1.4 billion in
2018~\cite{doi:10.18111/9789284421152}
and the ``flat-rating'' or elimination of international roaming charges~\cite{lh_regulation,roaming_charge,sheehy:nerdwallet} has led to 
an exponential growth in roaming traffic, expected to increase 32-times by 2022. 
At the same time, users' QoE expectations -- when using VoIP or posting videos -- has forced 
content and service providers to peer close to their users, wherever they may roam, thus adding 
to a growing interconnection ecosystem.

%Beyond support for people roaming, the \ac{IPX} Network with its global footprint has also become a catalyst 
%for cellular \ac{IoT}. MNOs' infrastructures offer the basic technological support for \ac{IoT} and \ac{M2M} services.
%Major \acp{MNO} exploit roaming and the IPX Network to give global connectivity to IoT providers, which ship their
%devices internationally (from wearables to cars and containers) with pre-arranged cellular service.\footnote{In contrast
%to an approach where IoT providers make local arrangements to obtain connectivity in \textit{each} country where their devices
%operate.}

\begin{figure}[!t]
	\includegraphics[width=.85\columnwidth]{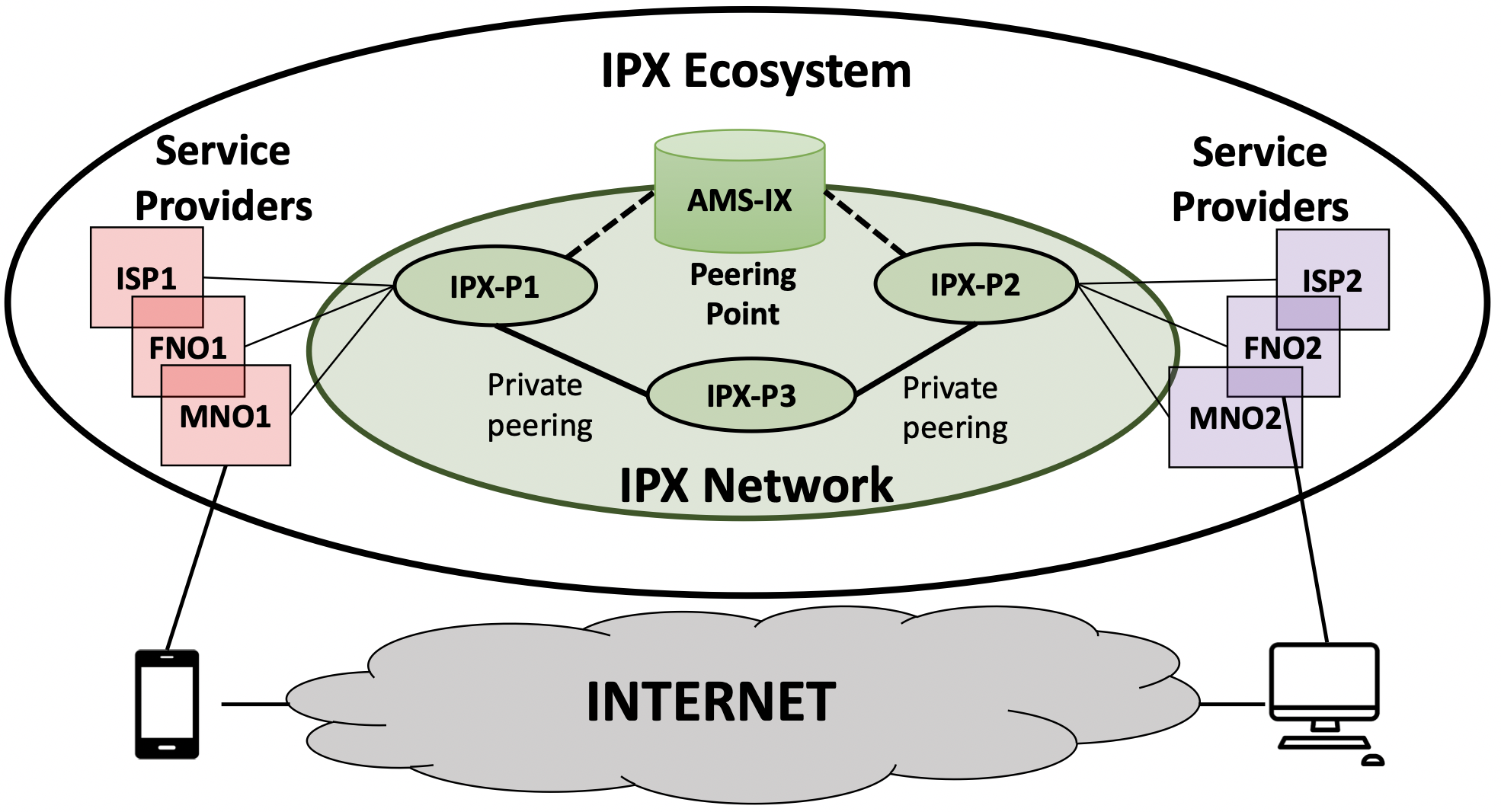}
	\caption{\small High level architecture of the IPX Ecosystem.}
	\label{fig:ipx_ecosystem}
	\vspace{-5mm}
\end{figure}

Despite its rapid growth and increased importance as the core of the mobile Internet, the IPX Network and its
associated ecosystem (Figure~\ref{fig:ipx_ecosystem}) has received little to no attention by our community, due in part to its intrinsic opacity and
separation from the public Internet. \textit{In this paper, we present the first characterization of the IPX ecosystem 
and a first-of-its-kind, detailed study of a large operational IPX Provider.}

We contribute the first topology analysis of the IPX ecosystem.  
We analyze a private BGP routing table snapshot from an operational router that is part of the IPX Network to map
the interconnection between \acp{IPX-P} and \acp{SP} for the data roaming service (\S~\ref{sec:ipx_ecosystem}).
We build the exhaustive list of 29 active IPX-Ps and detail their approach for peering using three major peering
exchange points (AMS-IX Amsterdam, Equinix Ashburn and Equinix Singapore).
\footnote{IPX is not the same as IXP, though the acronyms are similar. IPX-Ps may rely on IXPs (such as AMS-IX) for peering with other IPX-Ps.}
We capture the breadth and full-mesh peering fabric of the IPX Network for data roaming, which enables inter-working between all $\approx800$ \acp{MNO} currently active world-wide. 

We present a first-of-its-kind detailed analysis of a large operational IPX-P system, and provide the insider view into the otherwise inaccessible IPX Network.
Our study is based on real-world traffic records for its main service (radio signaling for data roaming (\S~\ref{sec:characteristics})) over a two-week period in December 2019. 
We showcase the operational IPX-P's signaling and data roaming infrastructures (\S~\ref{sec:signaling_analysis}), with notable presence in Europe and the Americas.
We study signaling traffic patterns (for different radio access technology) and data communications from over 22 million mobile devices roaming in the world.

% !TeX root = sigcomm20.tex
% ================================================================

\section{The IP eXchange Model}
\label{sec:overview}

In this section, we provide a detailed description of IPX and use data roaming, one of the main services 
offered, to illustrate the main IPX players and their interactions.

%%****************************************

\subsection{IPX Connectivity for SPs}
\label{sec:what_isNot_IPX}

At a high-level, the IPX ecosystem (Figure~\ref{fig:ipx_ecosystem}) includes Service Providers (SP) and 
networked IPX-Providers (IPX-Ps). 
\acp{IPX-P} are third-party interconnection providers to \acp{SP} (e.g., \acp{MNO}, \ac{IoT} providers).
\acp{IPX-P} peer with other \acp{IPX-P} to extend their footprint worldwide. 

While IP-based, the resulting IPX Network is a private network, separate from the public Internet, that meshes together the infrastructures of the IPX-Ps.  
It guarantees traffic separation between IPX services from the rest of the Internet. The IPX Network 
enables the transport of global roaming data between networks, with interoperability of different implementations 
and standards. 

\acp{SP} require a single connection and agreement with one IPX-P in order to connect to the IPX Network,
and interconnect with partner \acp{SP} world-wide.\footnote{Although direct interconnection between \acp{SP}
through leased lines or \ac{VPN} is possible, it is outside the scope of our analysis.} 
For instance, to enable data roaming, two MNOs must each have an agreement with an IPX-P in order to interconnect.
For redundancy, a \ac{SP} could establish connections to more than one \ac{IPX-P}.
Depending on the footprint of the IPX-P's infrastructure, \acp{SP} can use one or more \acp{PoP} of the IPX-P.

%****************************************

\subsection{Study Case: Data Roaming over IPX}

%\begin{figure}[t!]
%	\centering
%	\begin{subfigure}{0.95\columnwidth}
%		\includegraphics[width=\linewidth]{figures/GRX_example.png}%
%		\caption{IPX example of business roaming agreements.}%
%		\label{fig:grx_example}%
%	\end{subfigure}%
%	
%	\begin{subfigure}{0.95\columnwidth}
%		\includegraphics[width=\linewidth]{figures/roaming_config_2.png}%
%		\caption{Roaming configurations over IPX.}%
%		\label{fig:roaming_type}%
%	\end{subfigure}%
%	
%	\caption{\small Example of \acp{MNO} using the IPX for data roaming.}
%	\vspace{-3mm}
%\end{figure}

To establish roaming, roaming partner \acp{MNO} 
must have a functioning commercial agreement, implement their roaming technical solutions, establish 
inter-working and deploy their billing function. %The next paragraphs discuss each of these steps in detail.
We use data roaming to illustrate the main IPX ecosystem players and their interactions. 

In terms of business agreement solutions, the legacy option for \acp{MNO} is a standard \emph{bilateral agreement} 
where the two parties involved define terms and conditions of their cooperation. These bilateral roaming agreements 
for roaming and inter-working are costly and generally of lower value today, something that has served as additional 
motivation for \acp{MNO} to adopt the IPX model.  % \enote{FEB}{Why "lower'' value?}

Under the IPX model, operators connect to an \ac{IPX-P} to gain access to many roaming partners world-wide, externalizing 
the inter-working establishment to the \ac{IPX-P} offering the service. \acp{IPX-P} are then peering with each-other to 
expand their international footprint through the IPX Network. This IPX {\it hubbing} solution 
does not preclude the existence of bilateral agreements between \acp{MNO}, which can be viewed as a complementary 
roaming model. 
%Figure~\ref{fig:grx_example} illustrate the inter-working of two \acp{MNO} (i.e., the \ac{HMNO} and the \ac{VMNO}) for data roaming. 

Once a commercial agreement has been created, the \ac{IPX-P} sets up the technical roaming solution, including 
coordination over the signaling platform, and establishes the \ac{IPX-P} interconnectivity. 
After \acp{MNO} establish roaming inter-working, they deploy the billing service, which is 
key to recovering roaming revenue.  The roaming partners must each record the activity of roaming users in a given 
\ac{VMNO}. Then, by exchanging and comparing these records, the \ac{VMNO} can claim revenue from the partner \ac{HMNO}.

When a mobile device is at home, the subscriber's traffic will take a short path inside the network to reach a suitable \ac{PGW} to the Internet. 
When inter-working exists between two \acp{MNO}, there are several network configurations the IPX Network supports to enable roaming. 
\textit{The IPX-P's main function is to build the communication tunnel between the \ac{SGW} and the \ac{PGW}, enabling traffic to flow to and from the roaming mobile device.}
The traffic of a roaming mobile device is directed to an egress \ac{PGW} whose location depends on the roaming configuration.  
%Figure~\ref{fig:roaming_type} illustrates a set of configurations for roaming over the IPX Network -- \ac{HR}, \ac{LBO} and \ac{IHBO}. 
Different configurations for roaming over the IPX Network are available -- \ac{HR}, \ac{LBO} and \ac{IHBO}. 
%In the case of \ac{HR}, the mobile device receives the IP address from its home \ac{MNO} and 
%the roaming traffic is then routed over a tunnel between the \ac{SGW} in the \ac{VMNO} and a \ac{PGW} in the \ac{HMNO}. 
%With \ac{LBO}, the mobile node receives its IP address from the \ac{VMNO} and the traffic is 
%routed towards a local \ac{PGW} in the \textit{visited network}. 
%When using \ac{IHBO}, the mobile device obtains its IP address from the IPX Network and the traffic is routed through a \ac{PGW} in the \ac{IPX} Network. 
Prior work found that the default roaming configuration majority \acp{MNO} currently use in Europe is the \ac{HR} roaming~\cite{mandalari2018experience}. 
In the case of \ac{HR}, the mobile device receives the IP address from its home \ac{MNO} and 
the roaming traffic is then routed over a tunnel between the \ac{SGW} in the \ac{VMNO} and a \ac{PGW} in the \ac{HMNO}. 

% definitions and overview 

% !TeX root = sigcomm20.tex
% ================================================================

\section{IPX Ecosystem Diversity}
\label{sec:ipx_ecosystem}

The core players of the IPX ecosystem are \acp{IPX-P} and \acp{SP}. \acp{IPX-P} provide the interconnection 
between SPs directly through their network or via peering with other \acp{IPX-P} (see Figure~\ref{fig:ipx_ecosystem}).
The resulting ecosystem has a layered topology: a core of tightly interconnected IPX-Ps in a full mesh (the IPX Network) and the edge of diverse SPs that interconnect through the IPX Network, either through a single IPX-P or multi-homed through multiple IPX-Ps. 
For this, SPs may use their own access network (e.g., fixed and mobile network operators) or use a local provider to connect to the \ac{PoP} of the IPX-P (e.g., \ac{ASP}).
\acp{SP} originate and/or terminate traffic for one or several services; they do not transport traffic.

\subsection{IPX-Ps Interconnection and Peering}

IPX-Ps establish interconnection either through private bilateral interconnections or through an \ac{IXP} (Figure~\ref{fig:ipx_ecosystem}). 
The benefits of peering are well known among \acp{ISP} and \acp{CDN}, particularly when it comes to public peering via an \ac{IXP}~\cite{ager2012anatomy}.  
In recent years, we have seen more efforts to expand the peering culture to the mobile ecosystem ~\cite{apnic17peering}, by establishing new mobile peering infrastructure world-wide. 
This is (slowly) propagating to the mobile industry, and the interconnection of \acp{IPX-P} via mobile peering at specific \acp{IXP}, which offer this service, is becoming commonplace. 
Currently, the two major \acp{IXP} offering the mobile peering service are AMS-IX and Equinix, with three locations overall (Amsterdam, Ashburn and Singapore). 
 
IPX-Ps dynamically exchange routing information with other IPX-Ps using the BGP routing protocol. 
According to GSMA reccomendations~\cite{ir34}, IPX-Ps should not act as a transit IPX-P (i.e., there can only be a maximum of two IPX-Ps between two partner SPs). 
Therefore, when an IPX-P has a customer \ac{SP} who requires a connection to a customer \ac{SP} of another IPX-P, the two IPX-Ps should peer, either through (direct) private peering or peering points.
For example, in Figure~\ref{fig:ipx_ecosystem}, the path between MNO1 and MNO2 should only traverse two IPx-Ps (i.e., MNO1 -> IPX-P1 -> IPX-P2 -> MNO2). 
Also, IPX-P3 should never transit traffic for neither of its two peers, IPX-P1 and IPX-P2. 
To ensure this, network routes IPX-P3 receives either over private peering or over a peering point should not be re-advertised to other IPX-P peering partners. 
These recommendations result in a tightly interconnected IPX Network, with a theoretical diameter of two entities between any pair of SPs. 

\begin{figure}[!tbh]
	\centering
	\begin{subfigure}{0.5\columnwidth}
	\includegraphics[width=\linewidth]{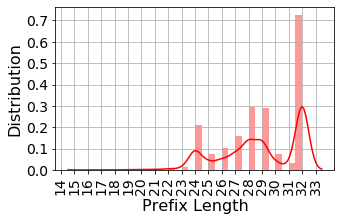}%
	\caption{IPv4 prefix length.}%
	\label{fig:pref_len}%
	\end{subfigure}%
    \begin{subfigure}{0.5\columnwidth}
	\includegraphics[width=\linewidth]{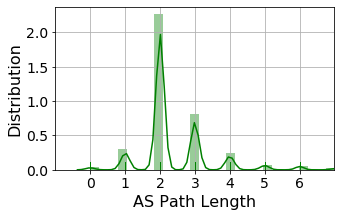}%
	\caption{AS-Path length. }%
	\label{fig:as_path_len}%
	\end{subfigure}%	
	\caption{\small Analysis of the prefixes within the routing table snapshot from an operational IPX-P: (a) distribution on prefix length; (b) distribution on AS-Path length (without AS-Path prepending).}
	\label{fig:prefixes_path}
	\vspace{-3mm}
\end{figure}

\subsection{IPX Topology}
\label{sec:snapshot_ecosystem}

Given the opaque nature of the IPX ecosystem, we cannot capture its characteristics from the public Internet, not by using public Internet routing data nor with active end-to-end measurements (e.g., traceroute). 
In order to shed light on the interconnection fabric between IPX-Ps and SPs in the IPX ecosystem, we analyze three different datasets: 
(i) \textit{routing dataset:} a private BGP routing table snapshot for data roaming that one of the largest operational IPX-Ps provided us;  
(ii) \textit{peering dataset:} the list of AMS-IX members that connect to the mobile peering services for data roaming service only,
together with the full internal list of IPX-P peers from the operational IPX-P providing us the routing dataset;
(iii) \textit{survey dataset:} market surveys and reports~\cite{rocco, hot_telecom} from third parties. 
In the routing data, we capture one snapshot of the routing table on the 30th of January 2020, which provides one view of the IPX-P's relationships for the data roaming service. 
We note that this information, though descriptive of the ecosystem, might be incomplete (i.e., there might be information not included here, but present in snapshots at other vantage points). 

Specifically, the IPX-P routing table snapshot includes reachability information for a total of 10,418 IPv4 prefixes advertised by 59 different entities (neighbors), which include the peer IPX-Ps and the customer MNOs of the IPX-P we analyze.   
By checking the originating AS of the prefixes in the routing dataset, we find that within the IPX ecosystem there are a total of 824 different service providers for data roaming (i.e., MNOs, MVNOs or M2M platforms).
%Given the previous recommendation of full-mesh interconnection between IPX-Ps, we conclude that this represents all the MNOs active world-wide. 
This number is consistent with the total number of \acp{MNO} active world-wide that register with the GSMA. 
These are public IPv4 prefixes that IPX-Ps do not announce in the global BGP routing tables.
Hence, they are not reachable from the public Internet. 
From Figure~\ref{fig:pref_len}, we note that the median prefix length in the routing dataset is /29, and there is a large number of /32 prefixes that MNOs originate.
These likely represent specific elements (e.g., \ac{HLR} or \ac{MME}) within the MNO infrastructure, which are involved in procedures for data roaming.

\textbf{IPX Network: }
We merge and corroborate the information we extract from the above-mentioned datasets to build a list of IPX-Ps that currently form the IPX Network.  
In Annex~\ref{annex_a}, we detail the full list of these providers and the methodology we found we compile this list. 
We specifically note that, in light of the growing popularity and worldwide footprint of IPX services, many heavyweight 
telecoms are participating in the IPX environment, leveraging their underlying extensive infrastructure.

We verify that these IPX-Ps appear in the routing dataset as active peers of the IPX-P providing us the routing table snapshot. 
The total number of entities advertising reachability information to the IPX-P is equal to 59 different ASes. 
Out of these, we separate the ones that advertise prefixes with an AS-Path length longer than one (i.e., they advertise their SPs).
We check the overlap with the list of IPX-Ps we built, and find a set of 23 different ASes.\footnote{The remaining 36 ASes observed in the BGP routing table are SPs that are customers of the analyzed IPX-P.}

\begin{figure}[tbh]
	\centering
	\begin{subfigure}{0.5\columnwidth}
		\includegraphics[width=\linewidth]{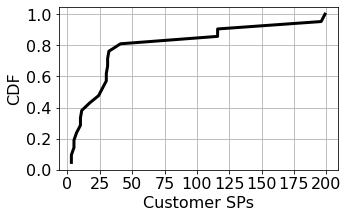}%
		\caption{\tiny Customer SPs per IPX-Ps. }%
		\label{fig:customers}%
	\end{subfigure}%
    \begin{subfigure}{0.5\columnwidth}
    \includegraphics[width=\linewidth]{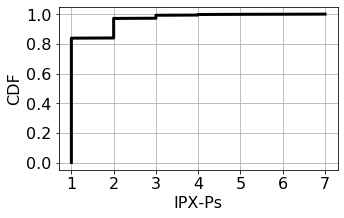}%
	\caption{\tiny IPX Providers per SP.}%
	\label{fig:multihomed}%
	\end{subfigure}%	
	\caption{\small Interconnection between IPX-Ps and SPs: (a) distribution of number of customers per IPX-P; (b) distribution of number of IPX providers per SPs. }
	\label{fig:interconnections_sp_ipxp}
	\vspace{-3mm}
\end{figure}

\textbf{IPX Interconnections: }
The set of 29 IPX-Ps we found must provide interconnection services to the over 800+ \acp{MNO} currently active world-wide~\cite{GSMA-MNOs}.
%\footnote{https://www.gsma.com/membership/membership-types/operator-membership/}
The number of \acp{PoP} indicates the number of locations where an IPX-P can cross-connect with \acp{SP}, giving insight into its world-wide geographical footprint. 
For example, large players such as Telia Sonera, TATA Communications, Orange, Vodafone, 
Telecom Italia, Telefonica or Telekom Austria offer an infrastructure with more than 100 \acp{PoP} world-wide each. 
The average number of PoPs among the 18 IPXs publicly disclosing this information is 116, hinting the breadth of the system we capture. 

We further use the routing dataset in order to characterize the interconnection between IPX-Ps and SPs in the ecosystem. 
For the IPX-Ps present in the routing dataset, we verify the different number of SPs for which they advertise reachability information (Figure~\ref{fig:customers}).

\textit{We find that four major players within the IPX Network (namely, Syniverse, BICS, Orange and Comfone) together give services to a total of more than 600 MNOs (out of the total 800). }
Inversely, we also verify the popularity of multi-homing among SPs. In other words, we quantify the number of IPX-Ps that advertise reachability information for the same SP. 
Figure~\ref{fig:multihomed} shows that 80\% of SPs are single-homed (i.e., they only connect to one IPX-P), while for the rest we observe up to seven different IPX-Ps. 
In particular, we note that multiple M2M platform providers use at least four different IPX-Ps, which is intuitive due to their reliance on roaming. 

Finally, in order to characterize the interconnection between IPX-Ps, we analyze the AS-Path of the prefixes in the routing dataset. 
To comply with the recommendation of a fully connected IPX Network, there can be no more than two different IPX-Ps involved in the communication between two different SPs. 
Figure~\ref{fig:as_path_len} shows the distribution of prefixes on AS-Path length. 
Note that we eliminated AS-Path prepending (used by IPX-Ps and SPs for traffic engineering).

\textit{We show that, indeed, majority prefixes advertised by peer IPX-Ps have a path length equal to two, confirming the tight interconnection required in the IPX Network.}
%This confirms that through tight interconnection in the IPX Network, the IPX-P is able to increase the footprint of its customer MNOs world-wide. 
The paths longer than two ASes represent the result of MNOs working together with their parent networks (e.g., national MNOs connecting to their parent carrier) or third party network providers (e.g., for MVNOs, M2M platforms) to achieve a broader geographical footprint.
Thus, sibling ASes that belong to the same organization appear in the same AS-Path. 

We also observe that ASes in the IPX network use standard techniques for traffic engineering. In particular, we observe a heavy use of AS prepending. AS path prepending makes a route less preferred to receive traffic by making the AS path length for the route artificially long by repeating ASes in the AS path attribute. We observe 1,331 routes from 131 different origin ASes where prepending was used. We also observe extensive use of Multi Exit Discriminator (MED), a BGP attribute that serves to express preference between different links between two ASes. We observed 1,583 prefixes that contained the MED attribute.
	
% description of the whole ecosystem 

% !TeX root = sigcomm20.tex
% ================================================================

\section{A Large IPX Provider}
\label{sec:characteristics}

We continue our analysis of the IPX ecosystem and zoom into the operations of one of the largest \acp{IPX-P} 
in the ecosystem. 
%We describe next the IPX-P's infrastructure and services, as well as the monitoring datasets we collect, which enable us to present a detailed view of the IPX-P's international footprint and operations. 
%To the best of our knowledge, this is the first analysis of the operational system of an IPX-P.
The IPX-P we dissect is a Tier-1 Internet Service Provider operating one of the 
largest backbone networks world-wide. As part of its interconnectivity products, the carrier operates 
an IPX infrastructure that runs on top of its vast \ac{MPLS} network.\footnote{An IPX-P requires access to an 
underlying backbone network. The \ac{IPX-P} may own its own \ac{MPLS} network or alternatively, it 
might lease capacity on MPLS networks on which they deployed the infrastructure needed to
deliver and manage inter-operable cross-network services.} 
\textit{The IPX-P infrastructure integrates 
more than 100 \acp{PoP} in 40+ countries with a particularly strong presence in America and Europe.  
}

%The services that the IPX-P supports include a wide selection across all the four layers that form the general
%IPX range of services, including IPX Transport, Data Roaming, SCCP 
%Global Signaling, LTE Diameter Exchange, \ac{M2M} service and other roaming value added services (e.g., 
%Steering of Roaming, welcome SMS, sponsored roaming, Data and Financial Clearing). 
	
In terms of network connectivity, the \ac{IPX-P} offers two types of interfaces, namely the IPX Access for 
clients (service providers) and the IPX Exchange for peering with other \acp{IPX-P}. The main mobile peering 
points the \ac{IPX-P} uses are those in Singapore, Ashburn and Amsterdam. By peering with other large Tier-1 
carriers, the \ac{IPX-P} extends its footprint world-wide to geographic regions where it does not own 
infrastructure (\S~\ref{sec:snapshot_ecosystem}). The \ac{IPX-P} serves clients in multiple countries in Europe (Germany, Spain, UK) and 
the Americas (including US, Mexico, Brazil, Argentina, Colombia, Peru, Chile, and Ecuador). 

%****************************************
\subsection{IPX-P Infrastructure and Monitoring}

We now describe the \ac{IPX-P} infrastructure, the monitoring methodology and the datasets we collected 
to characterize its operational system and services. 
We monitor the \ac{IPX-P} infrastructure corresponding to the two main services -- SCCP Global Signaling, LTE 
Diameter Exchange -- for two weeks, from December 1st to December 14th 2019.

\textit{Overall, the total number of IMSIs we capture is of more than 22M active daily in 2G/3G and more than 2M active daily in 4G/LTE.}\footnote{Note that there might be an overlap between these two sets. However, we aim to show here the load on the two different signaling infrastructures}. \textit{They correspond to 215 home countries and 210 visited countries. The \ac{IPX-P}'s customers are active within 19 countries and include \acp{MNO}, IoT/M2M connectivity providers and cloud service providers. }

\begin{figure}[t!]
  \includegraphics[width=.85\columnwidth]{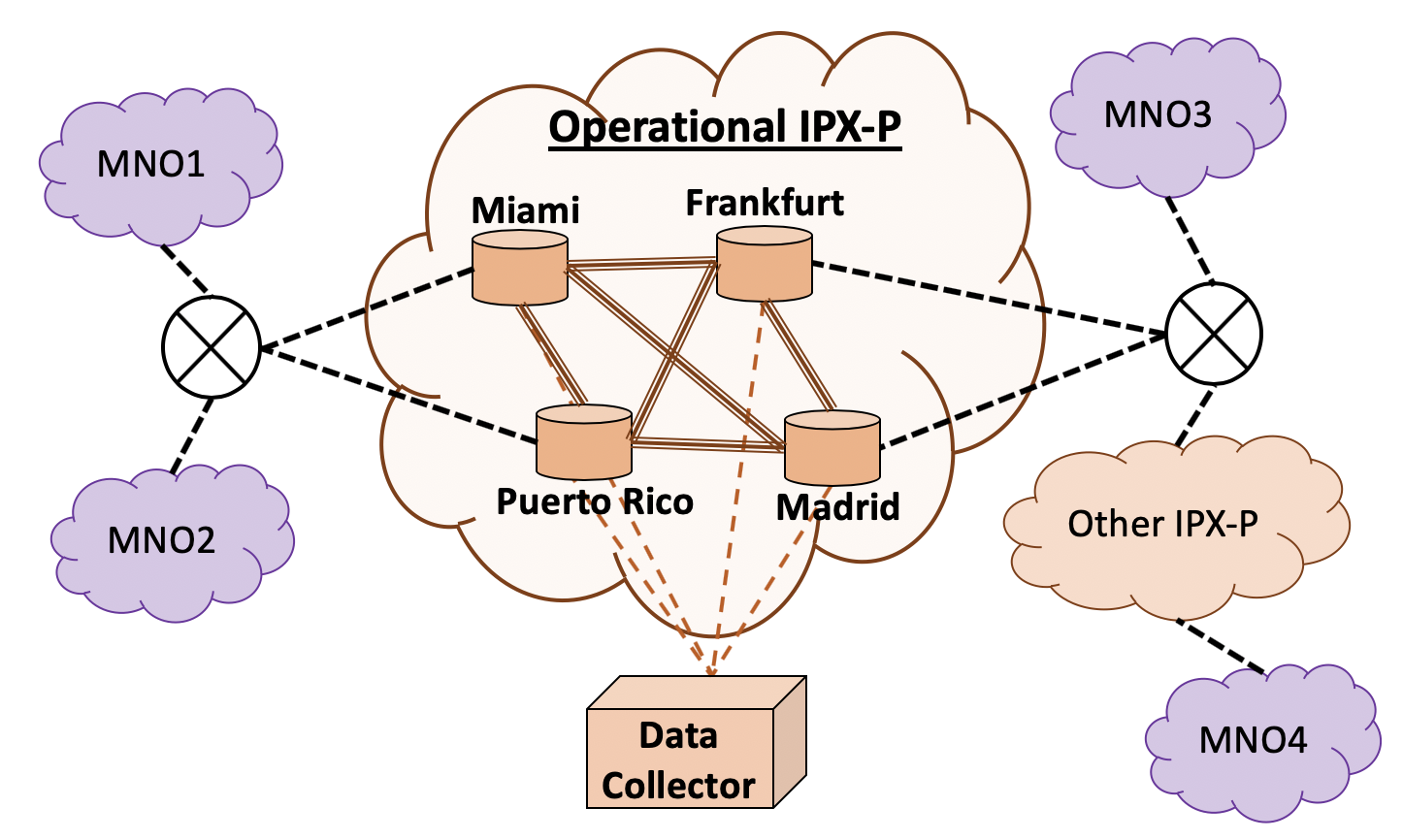}
   \vspace{-3mm}
  \caption{\small High level architecture of the IPX-P's signaling platform.}
  \label{fig:diameter_signaling}
  \vspace{-5mm}
\end{figure}

\paragraph{SCCP Global Signaling:}  This service provides access to the \ac{IPX-P}'s SS7 signaling network, satisfying the 2G/3G interconnection needs for international roaming of \acp{MNO}. The SCCP Signaling network of this particular IPX-P has a redundant configuration with four international \acp{STP} located in North America (Miami, Puerto Rico) and Europe (Frankfurt, Madrid) as depicted in  Figure~\ref{fig:diameter_signaling}. 
%The network enables access to over 1,000 destination world-wide either directly through the \ac{IPX-P}'s network or through peering agreements with other Tier-1 carriers (\S~\ref{sec:ipx_ecosystem}).

To capture clients' activity across this signaling platform, we monitor the \ac{MAP} protocol, which supports end-user mobility and is used by devices to communicate with the major network elements, including the \ac{HLR}, \ac{VLR} or the \ac{MSC}. %Figure~\ref{fig:diameter_signaling} shows the architecture of the SCCP platform and the location of the monitoring software allowing us to build the dataset we use in this paper. 
We collect traffic corresponding to the following procedures of each device belonging to one of the \ac{IPX-P}'s clients (outbound roaming) or to foreign devices that connect to the network of one of the \ac{IPX-P}'s clients (inbound roaming):  i) location management (update location, update GPRS location, cancel location, purge mobile device); ii)  authentication and security (send authentication information); iii) fault recovery. 

\paragraph{LTE Diameter Exchange:} This service provides the Diameter signaling capabilities necessary to enable 4G roaming for customers. 
The infrastructure of this particular \ac{IPX-P} includes four \acp{DRA} meant to forward Diameter messages and simplify interworking between different network elements. %It is application-unaware and does not inspect the messages it receives. 
%The service also integrates \acp{DPA}, which include the functionality of the \acp{DRA} and can additionally inspect and route Diameter messages based on different parameters. 
%Finally, by leveraging the Hosted \ac{DEA} service, the \ac{IPX-P} offers a infrastructure-as-a-service functionality to help operators expedite the launch of LTE roaming services. Thus, operators can use the dedicated customer virtual \ac{DEA} from the IPX-P instead of deploying their own infrastructure. 
The LTE Diameter service integrates value added services, including Welcome SMS, Steering of Roaming or Sponsored Roaming.

%The LTE Diameter Exchange service offers worldwide coverage to over 500 destinations directly connected to the \ac{IPX-P}'s network or reachable through agreements with peer \acp{IPX-P}. 
To monitor the activity of the \ac{IPX-P}'s customers across this platform, we monitor traffic across the geo-redundant signaling network with four \acp{DRA} located two in Europe (Frankfurt, Madrid) and two in North America (Miami, Boca Raton).
The infrastructure is similar to the one in Fig.~\ref{fig:diameter_signaling}. 

\begin{figure*}[th!]
	\centering
	\begin{subfigure}{0.33\textwidth}
		\includegraphics[width=\linewidth]{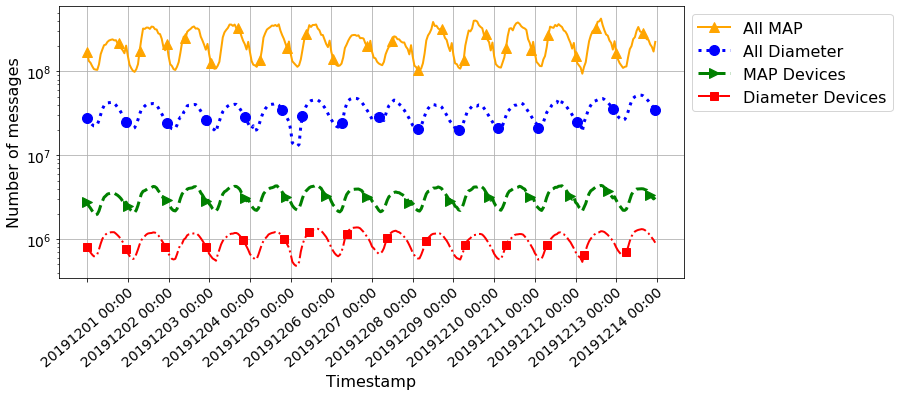}%
		\caption{\tiny Signaling traffic volume and number of devices in each platform (SCCP Signaling and Diameter
			Signaling).}%
		\label{fig:signaling_time_series_all}%
	\end{subfigure}%
	\begin{subfigure}{0.33\textwidth}
		\includegraphics[width=\linewidth]{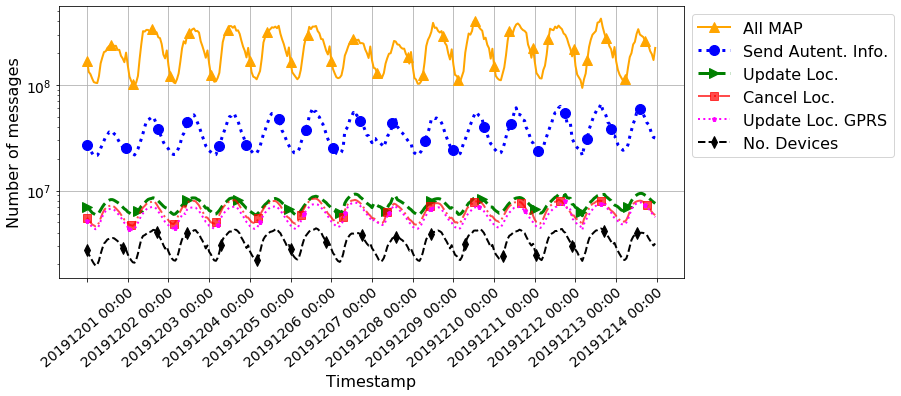}%
		\caption{\tiny SCCP signaling traffic time series; breakdown per type of signalling procedure.}%
		\label{fig:signaling_time_series_MAP}%
	\end{subfigure}%
	\begin{subfigure}{0.33\textwidth}
		\includegraphics[width=\linewidth]{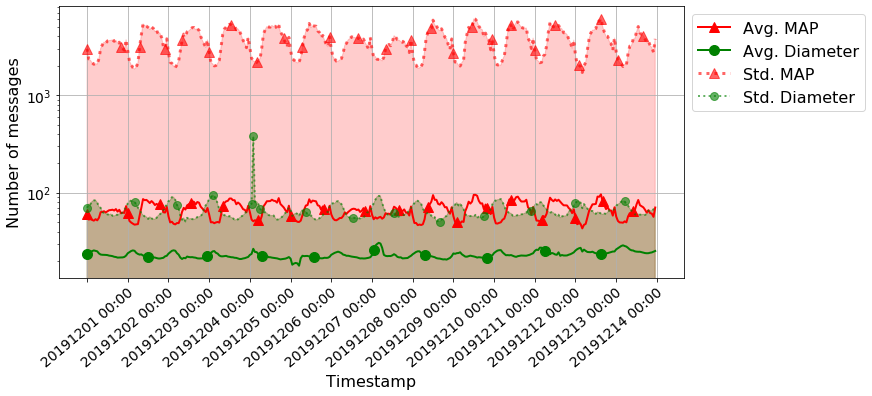}%
		\caption{\tiny Average and standard deviation of the number of SCCP messages and Diameter messages per IMSI per hour.}%
		\label{fig:signaling_avg_time_series}%
	\end{subfigure}%	
	\caption{\small Signaling traffic time series for the observation period of December 2019.}
	\label{fig:signaling_time_series}
	\vspace{-5mm}
\end{figure*}

\subsection{Signaling Traffic Trends}
\label{sec:signaling_analysis}

Figure~\ref{fig:signaling_time_series} shows signaling activity of roaming mobile subscribers during the 
observation period of December 2019. We look at both \ac{MAP} and Diameter signaling procedures. \ac{MAP} is the most 
important application protocol in the \ac{SS7} stack, and handles the roamers' mobility between countries. Although 
this is still the most used protocol for mobile interconnection application messages, the more recent 
Diameter~\cite{rfc6733} signaling protocol has been growing with the adoption of LTE.
Figure~\ref{fig:signaling_time_series_all} focuses on the total signaling traffic and on the number of different 
mobile subscriber devices that generate this traffic. 

\textit{We find that the number of devices using 2G/3G infrastructure 
(extracted from MAP traffic) is an order of magnitude higher than those using 4G infrastructure (based on the Diameter 
traffic). The volume of signaling traffic in the SCCP infrastructure is, correspondingly, more significant in terms 
of total volume than in the Diameter infrastructure.} 
We also note the typical daily and weekly traffic patterns on mobile subscribers' activities. For instance, December 1st was a Sunday, showing the expected decreasing trend in signaling traffic activity which can be seen again the following weekend (December 7-8th). 

Each record in this dataset represents a signaling procedure that a network element triggers, corresponding to different
standard routines. For instance, from the \ac{MAP} interface we capture mobility management routines, including 
location management and authentication. Figure~\ref{fig:signaling_time_series_MAP} shows the time series of signaling 
traffic broken down by type of signaling procedure, including Update Location (UL), Cancel Location (CL) and \ac{SAI} messages. 
The latter, \ac{SAI}, represents the highest fraction of MAP signaling traffic. 
Indeed, according to the GSM standard definition, the \ac{SGSN} in the visited network triggers the authentication of subscriber 
procedure upon IMSI attach, location update or before starting data communication, thus explaining the large volume of \ac{SAI} messages. 

% \ael{cite https://www.etsi.org/deliver/etsi_tr/123900_123999/123909/03.00.00_60/tr_123909v030000p.pdf}, 

Figure~\ref{fig:signaling_avg_time_series} shows the average number of records per IMSI calculated over all the IMSIs we
observe in each one-hour interval (continuous line) during the observation period, as well as the standard
deviation of the number of records per IMSI calculated over all the IMSIs active in the same one hour interval (shaded
area). We observe both the MAP procedures for 2G/3G (red color) and the similar Diameter procedures for 4G/LTE (green color). 
While Diameter and MAP are different protocols, the underlying functional requirements (e.g.,
authenticating the user to set up a data communication) have many similarities in terms of the messages used for
Diameter and the SS7 MAP protocol implementation. We note that the load in terms of average signaling records per IMSI is in
the same order of magnitude (the continuous lines on the plot), regardless of the infrastructure the devices use; yet, 
there are significantly more messages for MAP, as Diameter is a more efficient protocol than MAP~\cite{rfc6733, rfc7075}.
%as the MAP procedures are "chattier" than Diameter ones~\cite{rfc6733, rfc7075}.

\begin{figure}[tb!]
	\centering
	\begin{subfigure}{0.5\columnwidth}
		\includegraphics[width=\linewidth]{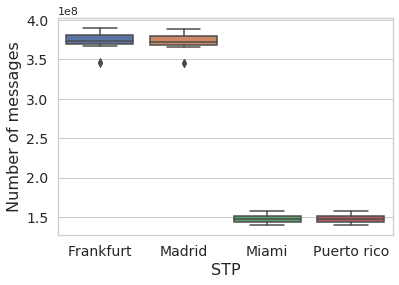}%
		\caption{\tiny  SCCP global signaling.}%
		\label{fig:sig_load_map}%
	\end{subfigure}%
	\begin{subfigure}{0.5\columnwidth}
		\includegraphics[width=\linewidth]{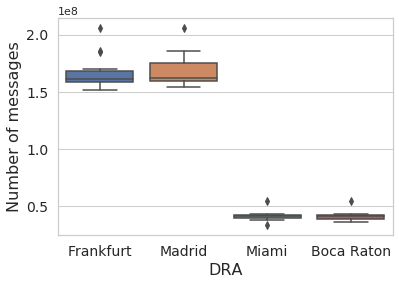}%
		\caption{\tiny Diameter signaling.}%
		\label{fig:sig_load_diameter}%
	\end{subfigure}%
	\caption{\small Signaling traffic load (total number of records) per infrastructure element per day for SCCP platform
		and Diameter. The boxplots capture the first two weeks in December 2019. }
	\label{fig:sig_load_infrastructure}
	\vspace{-5mm}
\end{figure}

We further investigate the traffic load on the signaling infrastructure, both for the SCCP (with \acp{STP} in Frankfurt,
Madrid, Miami and Puerto Rico) and the Diameter signaling infrastructure (using \acp{DRA} in Frankfurt, Madrid, Miami
and Boca Raton). 
Figure~\ref{fig:sig_load_infrastructure} shows the aggregated signaling traffic (number of messages) per infrastructure
point per day over the same observation period. We note that the redundant deployment of infrastructure
points in each geographical area allows the IPX-P to load-balance the signaling traffic across them (e.g., for 4G
between Frankfurt and Madrid in Europe and between Miami and Boca Raton in America). Signaling traffic in Europe through
the signaling points in Frankfurt and Madrid is considerably higher than that flowing through the signaling points in
America (Miami and Boca Raton/Puerto Rico). This is true for both the SCCP infrastructure
(Fig.~\ref{fig:sig_load_map}) and the Diameter infrastructure (Fig.~\ref{fig:sig_load_diameter}). When comparing the
two signaling services (i.e., SSCP and Diameter), we also note that the traffic volume in the SCCP signaling
infrastructure is approximately twice larger than the traffic volume flowing through the Diameter signaling
infrastructure. This proves the continued popularity of 2G/3G services in the region served by the IPX-P.

% !TeX root = sigcomm20.tex
% ================================================================

\section{Related work}
\label{sec:related}

The growing demand for global, mobile broadband access and a shift to all-IP-based services (from
broadband last mile to VoIP) have brought new impetus to the old idea of the IPX, first proposed by
the GSMA in 2007 to replace the traditional, bilateral-agreement model for international roaming~\cite{ir34}.
Despite the continuous technical development by IPX-Ps and the related parties ~\cite{US20140169286A1,
US20150222554, US20120218924A1} there has been few academic works on the topic. Takaaki~\cite{takaaki:survey} 
provides an early survey of IPX and its technical requirements, but there have been no in-depth analysis of IPX 
since due, in part,  to its closed nature. Our work presents the first in-depth analysis of the IPX Network 
and the associated ecosystem.

\section{Conclusion}
\label{sec:conclusion}

%The \ac{IPX} network interconnects 
%\acp{MNO} and a range of other \acp{SP} worldwide, enabling data roaming and supporting a variety of emerging applications. Recent years have brought a rapid growth in the number of participants in the IPX ecosystem and the volume of traffic 
%they exchanged. However, despite its importance, little is known about the IPX Network and IPX providers in the research community. 

In this paper, we provided both a qualitative and quantitative description of the IPX ecosystem using information collected from the one of the largest IPX providers. 
We believe that understanding the IPX network is cornerstone to understand and evolve the mobile Internet, and that it will become more relevant as new services emerge. 
For example, relying on IPX services, novel technologies such as eSIM that allow remote provisioning of mobile devices permit Global Mobile Network providers (e.g., Truphone) to emerge and offer novel connectivity options to end-users.

\balance
\bibliographystyle{ACM-Reference-Format}
\bibliography{reference}

\pagebreak
\appendix
\section{Ethical considerations}
Data collection and retention at network middle-boxes are in accordance with the terms and conditions of the \ac{IPX-P} and the local regulations, and only with the specific purpose of providing and managing the IPX service. 
The terms also include data processing for monitoring and reporting as allowed usages of collected data. 
Data processing only extracts aggregated information and we do we not have access to any personally identifiable information. 
We nevertheless consulted with the \ac{IRB} office at our institution who confirmed that no \ac{IRB} review was necessary as the study relies on the analysis of de-identified data. 

\section{IPX Providers}
\label{annex_a}

\textit{We compiled a list of 29 IPX-Ps that, at the time of writing, interconnect to form the IPX Network. }
Table~\ref{table:ipx-providers} lists the IPX-Ps we identified as active in the IPX Network, their (public) peering policy, number of \acp{PoP} 
and the type of \ac{AS} number (private/public). %, the number of IPX-P peers visible from the public Internet. 
We built the list of IPX-Ps and the information we show in the table by manually exploring the peering dataset. 
Specifically, at AMS-IX we found 27 customers for the mobile peering service with IPX/GRX tags~\cite{amsix}. 
In addition, we found 24 of the total 27 IPX-Ps peering at AMS-IX also present at Equinix IXPs. 
According to GSMA, the majority of IPX-Ps connect to these two IXPs~\cite{ir34}. 
Additionally, we checked the top 10 largest IXPs and some global IXPs and found no additional 
IXPs offering the mobile peering service.
We complete this list with two additional IPX-Ps we observe in the internal list of peers from the operational IPX-P, thus bringing the total number of players in the IPX Network to 29, which we list in Table~\ref{table:ipx-providers}.
We confirm this list and the IPX-Ps' identities by checking several commercial market surveys and reports about IPX-Ps from diverse parties~\cite{hot_telecom, rocco, nanog51cowie}.

We also find that among the set of currently active IPX-Ps, there are several which focus on interconnecting SPs within a specific region (e.g., Telin Indonesia, SAP). We do not include these in our analysis, but instead focus on the list we show in Table~\ref{table:ipx-providers}. 

The number of \acp{PoP} in Table~\ref{table:ipx-providers} indicates the number of locations where an IPX-P can cross-connect with \acp{SP}, giving insight into its world-wide geographical footprint. 
The average number of PoPs among the 18 IPXs publicly disclosing this information is 116. 

\begin{table}[!thb]
	\small
	\caption{List of active IPX Providers.}
	\label{table:ipx-providers}
	\vspace{-2mm}
	\begin{tabular}{m{2.5cm}|m{0.7cm}|m{2.2cm}|p{0.6cm}}
	{\bf IPX-P Name} 					& {\bf BGP} 	& {\bf Peering Policy} 			& {\bf PoPs} \\
	\hline 
	BICS 										& X 			& case-by-case 			& 120+ \\ 
	BT 											&   			& IPX/GRX only 			& \\ 
	C \& W 									& X  			& closed 						& \\ 
	China Mobile 							&  			& custom 					& 9 \\ 
	CITIC Telecom 						&  			& closed 						& 20+ \\ 
	Comfone 								&  			& closed 						& 11 \\ 
	Deutsche Telekom 				& X 			& case-by-case 			& 30 \\ 
	Etisalat 									& X			& case-by-case 			& 14 \\ 
	HGC 										& X			&  								&  \\ 
	iBasis 										&  			&  								& ~100 \\ 
	MEO 										& X 			& open						&  \\ 
	MTT 										& X 			& open 						&  \\  % (CJSC Multiregional Transit Telecom)
	MTX hub									&   			& open 						&  \\  % 	MTX hub (MTX services Sarl)
	NTT 										& X  			& closed 						&  \\ 
	Orange 									& X 			&  								& 250 \\ 
	OTE Globe 								& X 			& open 						& 21 \\ 
	PCCW Global 							&  			& case-by-case 			&  \\ 
	Syniverse 								& X			& case-by-case 			&  \\ 
	TATA 										& X 			& case-by-case 			& 400+ \\ % TATA Communications
	TDC A/S 									& X 			& case-by-case 			&  \\ 
	Tele2 AB 								& X 			& selective 					&  \\ 
	Telecom Italia 						& X 			&  								& 122 \\ 
	Telefonica 								& X 			& case-by-case			& 100+ \\ 
	Telekom Austria 					& X 			&  								& 148 \\ 
	Telenor 									& X 			& closed 						& 13 \\ 
	TeliaSonera 							& X 			&  								& 300+ \\ 
	Telstra 									& X 			& open 						& 36 \\ 
	TNS  										& X			& open 						& 125 \\ % Transaction Network Services
	Vodafone 								& X 			& case-by-case 			& 273 \\ 
	\hline 
	\end{tabular} 
	\vspace{-3mm}
\end{table}

\end{document}